\documentclass[dvipsnames,aps,pre,nobibnotes,twocolumn,superscriptaddress]{revtex4-2}
\usepackage[T1]{fontenc}
\usepackage[english]{babel}
\usepackage[utf8]{inputenc}
\usepackage[unicode=true,pdfusetitle,bookmarks=true,bookmarksnumbered=false,bookmarksopen=false, breaklinks=false,pdfborder={0 0 0},pdfborderstyle={},backref=false,colorlinks=true]
 {hyperref}
 \hypersetup{
 citecolor=blue,
 urlcolor=black,
 linkcolor=black}
\usepackage{array}
\usepackage{units}
\usepackage{amsmath}
\usepackage{cancel}
\usepackage{mathrsfs}
\usepackage{amssymb}
\usepackage{graphicx}
\usepackage{stackrel}
\usepackage{babel}
\usepackage{xparse}
\usepackage{soul}
\usepackage[final]{pdfpages}
\usepackage{comment}
\usepackage{tcolorbox}
\usepackage{mathtools}

\makeatletter
\AtBeginDocument{\let\LS@rot\@undefined}
\makeatother

\begin{document}

\title{Fluctuating environments favor extreme reproductive delays and penalize intermediate ones}

\author{Jorge Hidalgo}
\affiliation{Departamento de F\'isica, Universidad de C\'ordoba, 14071 C\'ordoba, Spain}
\affiliation{Instituto Carlos I de F\'isica Te\'orica y Computacional, E-18071, Granada, Spain.}
\author{Lorenzo Fant}
\affiliation{Istituto Nazionale di Oceanografia e Geofisica Sperimentale (OGS), Trieste, Italia}
\affiliation{National Biodiversity Future Centre (NBFC), Palermo Piazza Marina 61, 90133 Palermo, Italy}
\author{Rafael Rubio de Casas}
\affiliation{Departamento de Ecolog\'ia, Universidad de Granada, Granada 18071, Spain}
\affiliation{Research Unit “Modeling Nature” MNat, Universidad de Granada, Granada 18071, Spain}
\author{Miguel A. Mu\~noz}
\affiliation{Departamento de Electromagnetismo y F\'isica de la Materia, Universidad de Granada, Granada 18071, Spain}
\affiliation{Instituto Carlos I de F\'isica Te\'orica y Computacional, E-18071, Granada, Spain.}
\date{\today}
          
\begin{abstract}
The timing of reproduction is crucial for biological fitness. Reproductive delays carry an inherent demographic cost: an individual that takes longer to reproduce contributes later to population growth, a priori, without any compensating benefit. 
However,  natural selection favours delayed reproduction under certain environmental conditions, a phenomenon that remains poorly understood. 
Here, we investigate how the environment can determine the evolution of particular life histories using a parsimonious delayed-logistic model. Individuals become reproductive only after a fixed lag whereas birth rates fluctuate under temporally correlated stochasticity.
Numerical integration, analytical calculations and evolutionary simulations reveal that the joint effect of demographic memory and colored multiplicative noise generate a strongly non-monotonic dependence of fitness on life history, with three distinct regimes of population performance: 
rapid reproduction maximizes linear growth but amplifies fluctuations and extinction risk; 
delayed reproduction buffers environmental variability, substantially increasing mean extinction times despite slower growth; 
strikingly, and central to our results, there is a broad band of maladaptive intermediate life histories that simultaneously reduce both growth and persistence, an effect that arises generically from the mismatch between reproductive timing and environmental autocorrelation. Our results show that reproductive timing is not merely a life-history parameter but an adaptive mechanism tuned to environmental timescales, and that “dangerous middle” maturation times can be inherently disfavored, with implications for understanding reproductive strategies across the tree of life. 
\end{abstract}
\maketitle   

\section{Introduction}
Life history theory addresses how organisms allocate time and resources among growth, reproduction, and survival \cite{Stearns1992,Levin1995}. A key component is the duration of the reproductive maturation period ---the interval from birth to first reproduction--- which carries a direct fitness cost: delaying reproduction reduces an individual’s contribution to population growth. Despite this, maturation times vary over orders of magnitude across taxa, from hours in microbes to decades in long-lived plants, and even among closely related species (e.g., annuals vs. trees in Euphorbia) \cite{HORN2012305,LeeOhSon2024EuphorbiaPlastome}. Explaining the evolutionarily optimal duration of this period remains a central challenge, particularly in fluctuating environments where reproductive timing strongly affects both growth and persistence.

Reproductive maturation reflects diverse biological processes that shape life stages. In some cases, its costs are offset by compensatory benefits: larger size at maturity can increase fecundity or competitive ability, and extended development may enhance adult survival \cite{Stearns1992}. However, such compensation is not universal. In many cases ---such as dormancy, which acts as a quiescent interval during which individuals neither grow nor reproduce--- its presence remains unclear.

Dormancy is widespread across bacteria, plants, fungi, insects, and even cancer cells \cite{Lennon2011,Baskin2,Denlinger}. It manifests itself in various forms, including microbial lag-time strategies tuned to environmental periodicity \cite{Fridman,Camacho}, synchronized reproductive cycles in insects \cite{Lloyd1966}, and broad germination-time distributions in plant seed banks under unpredictable rainfall \cite{Lennon2021,Lennon2017,Rees,Venable,Rubio2014,Lennon2023}. Despite this diversity, common patterns emerge: dormancy duration is closely linked to the duration and predictability of unfavorable conditions, and its timing is under strong selection. Mechanisms enabling environmental sensing—such as cue-driven germination in plants \cite{Donohue,Rubio2017}, diapause termination in insects and zooplankton \cite{Sweet2025,Wilsterman2020}, or exit from bacterial latency \cite{SachidanandhamGin2009Dormancy,McDonaldDormancy2024}—allow organisms to resume activity under favorable conditions. Similar dynamics appear in cancer, where dormant tumor cells reactivate in response to micro-environmental cues \cite{cancer1,cancer2,cancer-Wang}. At the same time, stochastic variation in dormancy duration is also common and can function as a bet-hedging strategy in unpredictable environments \cite{Cohen1966,Kussell-Leibler,Norman-review}. The coexistence of cue-driven responses and stochastic strategies highlights a complex and not yet fully resolved phenomenology.

Most theoretical frameworks justify reproductive delay through compensatory trade-offs \cite{ODwyer2023}, but whether such compensation is necessary—or even adaptive—remains unclear. This raises a fundamental question: can intrinsically costly delays be favored solely due to environmental fluctuations? More specifically, how does the temporal structure of environmental variability shape optimal maturation time, and can selection favor both precise environmental tuning and stochasticity? While prior work has examined timing strategies under periodic or uncorrelated fluctuations \cite{Hutchinson1948,Bulmer1984,Mitarai,Wide,powerlaw,Baranyi}, the role of temporal correlations has received less attention, despite strong empirical evidence that environmental variability is often autocorrelated (“colored” noise) \cite{Lewontin1969,tuljapurkar1990,Tuljapurkar}.

Here, we address these questions using a minimal delayed-logistic model under stochastic environmental conditions with finite correlation time. Individuals become reproductive after a fixed delay $\alpha$, introducing demographic memory, while birth rates fluctuate under temporally correlated noise. By isolating reproductive delay as a standalone trait, our framework complements models incorporating sensing or trade-offs and reveals how environmental correlations alone can shape the evolution of reproductive timing.

Our results uncover a rich and non-intuitive phenomenology. The interaction between delay and colored noise produces three dynamical regimes: short delays maximize mean growth but amplify fluctuations and extinction risk; long delays reduce variability and enhance persistence at the cost of slower growth; and, remarkably, intermediate delays perform worst, simultaneously lowering growth and resilience. This non-monotonic behavior arises from a mismatch between the delay timescale and environmental autocorrelation. Using an evolutionary agent-based model, we further show that this landscape leads to bistability, with populations evolving toward either short or long delays while avoiding the maladaptive intermediate regime. This pattern echoes empirical observations such as bimodal germination strategies in plants, extended dormancy in periodical cicadas, and mixed phenotypes in microbial persisters and zooplankton propagules \cite{Ratcliff,China,Cao,Berkens,Caceres}.

\begin{figure*}[t!]
    \centering
\includegraphics[width=\textwidth]{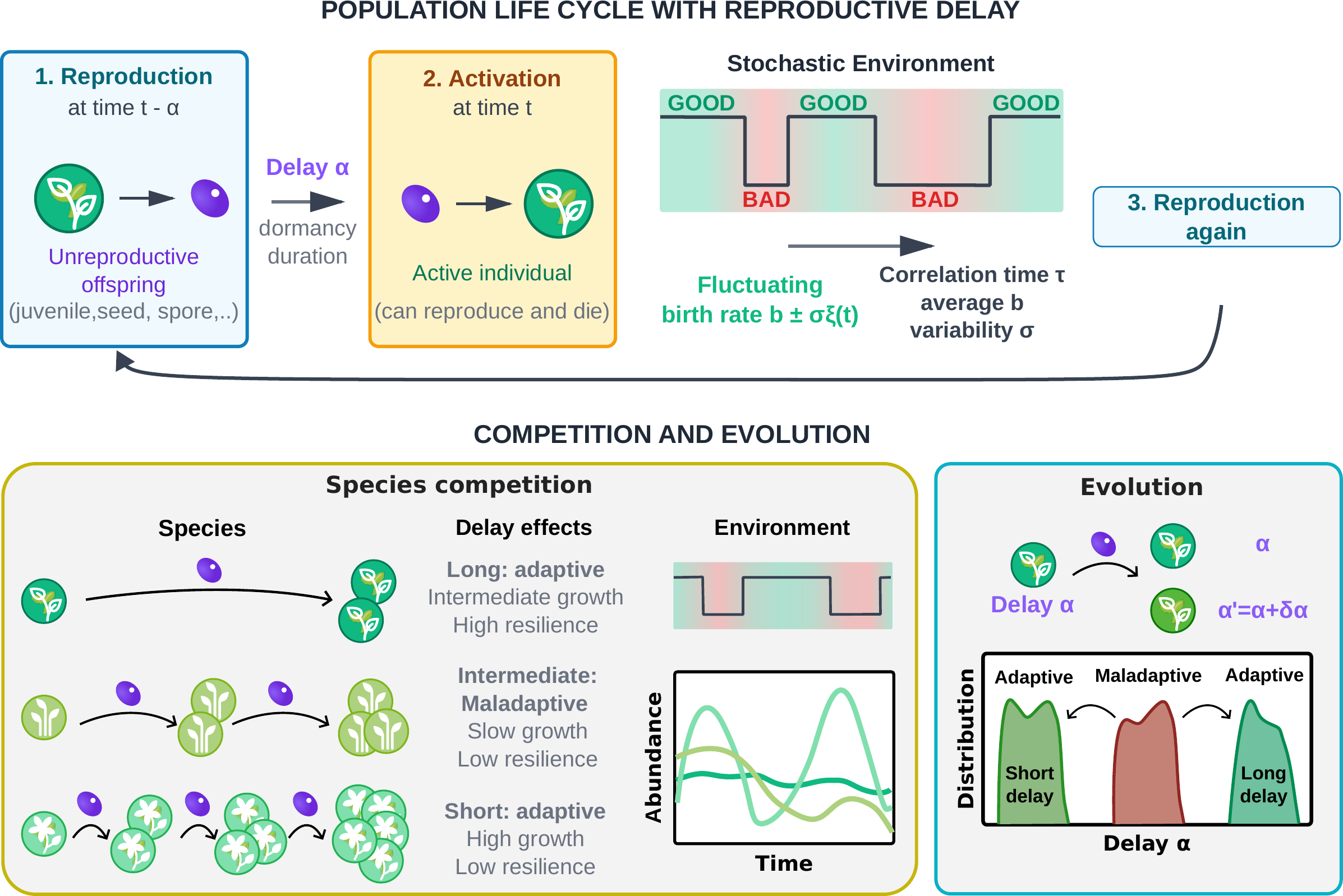}
\caption{Population life cycle with reproductive delay in a stochastic environment. Illustrated with plants, this framework can also be applied to other to organisms with maturation delay or dormancy. Individuals produce inactive propagules (e.g., seeds, spores, persister cells) at time $t - \alpha$, which remain in a quiescent state for a delay $\alpha$ before activating into reproductive, mortal individuals at time $t$. Active individuals experience alternating good and bad environmental conditions, represented as temporal fluctuations in the birth rate with mean $b$, amplitude $\sigma$ and correlation time $\tau$. We then use this framework to investigate (left) ecological competition among species that differ in reproductive delay and activation, and (right) evolutionary change in delay duration via single mutations that shift the delay from $\alpha$ to $\alpha' = \alpha + \delta\alpha$. In both cases, extreme strategies with either very long or very short delays gain a competitive advantage, whereas intermediate reproductive delay strategies are maladaptive and are progressively replaced by fitter ones.}
    \label{fig:scheme}
\end{figure*}

\section{Stochastic delayed-logistic model for populations with environmental noise and delayed reproduction}
\label{sec:mean-field-model}
Our model ---illustrated in Figure \ref{fig:scheme}--- considers a population undergoing logistic growth, governed by a differential equation that incorporates two key mechanisms: (i) environmental stochasticity and (ii) finite maturation time (interpreted as a reproductive delay) \cite{Hutchinson1948}
(see upper panel of Figure \ref{fig:scheme}).  In contrast to classical logistic models, reproduction is delayed and influenced by a correlated stochastic environment, while death takes place at a constant rate. In this formulation, the delay $\alpha$ represents the interval an individual must complete before it can reproduce. Consequently, the population density $x(t)$ evolves according to the following equation:
\begin{equation}
\dot{x}(t) = (b + \sigma \xi_\tau(t))\, x(t-\alpha) \left(1 - \frac{x(t)}{K} \right) - d\, x(t),
\label{eq:model}
\end{equation}
where $d$ is the per capita death rate (unless otherwise stated, we set $d=1$). Following reproduction at time $t-\alpha$, the offspring remains inactive/dormant for a fixed duration $\alpha$ and becomes active at time $t$, at which point it can reproduce or die. This leads to a delayed dependence in the birth term \cite{LaFuerza-Toral2013}, which is proportional to the population at the earlier time $t - \alpha$. Furthermore, the reproduction rate is modulated by the carrying capacity $K$ through a logistic saturation term $1 - x(t)/K$, evaluated at the time of becoming active. An extended formulation can incorporate $\alpha$ as a random variable drawn from a probability distribution, thus capturing individual-level variability in maturation times or dormancy: $x(t-\alpha) \rightarrow \int d\alpha\, \rho(\alpha) x(t-\alpha)$, where $\rho(\alpha)$ represents the distribution of reproductive delay durations across the population. Although this approach can describe more accurately many situations in nature ---including diversification and bet-hedging strategies (see below)--- for now we  treat $\alpha$ as a fixed control parameter.

Crucially, environmental variability is incorporated into the model through a time-dependent birth rate, which fluctuates as a result of external conditions \cite{color}. Specifically, the intrinsic birth rate is replaced by a stochastic term $b+\sigma \xi_\tau(t)$, where $\xi_\tau(t)$ represents environmental noise, with amplitude $\sigma$.

For simplicity, we represent environmental variability with a dichotomous continuous-time Markov noise (DMN)~\cite{Bena2006}, with states $\pm 1$ and symmetric transition rate $\tau^{-1}$. This modeling choice is mathematically tractable, maintains non-negativity of rates, and encodes alternating ``good'' and ``bad'' conditions. As a consequence, the process exhibits a temporal autocorrelation function $\langle \xi_\tau(t) \xi_\tau(s) \rangle = \exp(-2|t - s|/\tau)$, with characteristic correlation time $\tau/2$. We note that an alternative formulation where environmental variability affects the death rate instead of the birth rate is also possible (see Appendix~\ref{appendix:d}). 

The use of DMN ensures that the birth rate remains non-negative as long as $\sigma \leq b$, which is essential to prevent nonphysical scenarios in the presence of delay. In fact, if the coefficient multiplying the delayed term becomes negative, the dynamics can reach negative population densities \cite{kuang1993delay, Gros2019-reviewdelay}. For this reason, other forms of colored noise ---such as Ornstein--Uhlenbeck processes--- could induce negative fluctuations in the birth rate, potentially destabilizing the population. Nevertheless, we have verified that the qualitative behavior of the model remains robust under alternative noise models (see Appendix~\ref{appendix:OU}).

It should be noted that the environmental noise term $\xi_\tau(t)$ could, in principle, be evaluated either at the moment of reproduction $t-\alpha$, i.e. when the offspring is generated, or at the activation time $t$, i.e. when it becomes capable of reproduction and subject to death. In our case, this choice is irrelevant because of the temporal translational invariance of the DMN. However, this distinction may become relevant in extended models with stochastic $\alpha$ or with multiple interacting populations coupled to the same environmental fluctuations.

This modeling framework enables us to investigate the interplay between the reproductive delay ($\alpha$) and key features of environmental variability ---such as its amplitude ($\sigma$) and temporal correlation ($\tau$)--- and how their interaction shapes the long-term dynamics of populations exposed to fluctuating environments.

We integrate Eq.~\ref{eq:model} numerically using an explicit fourth-order Runge-Kutta (RK4) method adapted to handle delay differential equations, following the approach described in~\cite{Gros2019-reviewdelay} (see Appendix \ref{appendix:numericalintegration}). As is customary in delayed systems, initial conditions must be specified in the time  interval $[-\alpha,0]$. In our simulations, we set $x(t') = x(0)$ for $t'<0$. This choice has no qualitative impact on the long-term dynamics, since the presence of noise rapidly erases the memory of the initial condition. It should be noted that the most interesting phenomenology of the model emerges when the average reproduction rate is just above the critical threshold, $b \gtrsim d$, where the interplay between environmental stochasticity and delayed reproduction leads to rich dynamical behavior. In this marginal regime, fluctuations and memory effects can significantly impact population persistence and variability, making the role of reproductive delay particularly relevant (see Appendix \ref{appendix:b}). Such a regime is surely suited to describe populations in saturated environments where the overall growth rate is close to zero.
\begin{figure}
    \centering
    \includegraphics[width=\columnwidth]{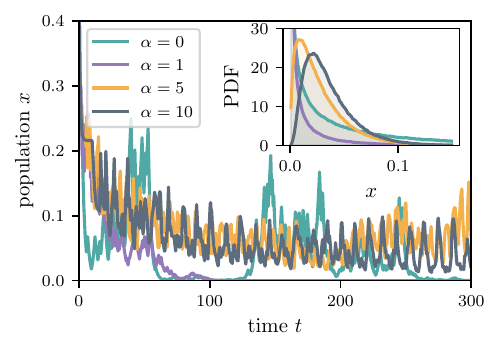}
    \caption{Time series of the population density $x(t)$ for different values of the reproductive delay $\alpha$, illustrating the effect of delayed reproduction under environmental noise. The inset shows the corresponding stationary probability density functions (PDFs) of $x(t)$, highlighting qualitative changes in population variability across delay regimes. Parameters: $b = 1.05$, $\tau = 1$, $\sigma = 0.5$, $d = 1$, $K=1$.}
    \label{fig:timeseries}
\end{figure}

Figure \ref{fig:timeseries} illustrates the results of computer simulations. In particular, it shows representative time series of the population density $x(t)$ for varying reproductive delay $\alpha$, under conditions of pronounced environmental stochasticity. In a nutshell, we identify three regimes: for short delays (e.g., $\alpha = 0$), the population grows rapidly but remains prone to extinction due to strong fluctuations (note that true extinction cannot occur in the absence of demographic noise \cite{Nature,Genovese}); at intermediate delays (e.g., $\alpha = 1$), the system consistently collapses to extinction. For large delays (e.g., $\alpha \gtrsim 5$), the population stabilizes and oscillates around a steady value, with reduced amplitude in both positive and negative fluctuations.

\subsection{Mean population diagram}

\begin{figure}
    \centering
    \includegraphics[width=\columnwidth]{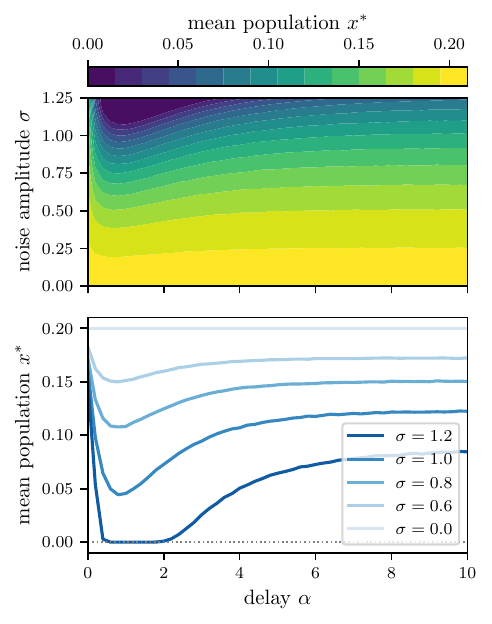}
\caption{
Mean population density $x^*$ as a function of reproductive delay $\alpha$ and noise amplitude $\sigma$. Upper panel: heat map of $x^*$ in the $(\alpha,\sigma)$ plane.  Lower panel: the same data shown as $x^*$ versus $\alpha$ for different fixed values of $\sigma$, highlighting its non-monotonic behavior.  To calculate $x^\ast$, we numerically integrate Eq.~\ref{eq:model} up to a burn-in time $T_{\min}$ to remove transients, compute the time average over the window $[T_{\min},\,T_{\max}]$, and then average over independent noise realizations. 
Intermediate delays combined with strong noise leads to a marked decrease in population size, reflecting poor timing strategies. Parameters: $d = 1$, $b = 1.25$, $\tau = 1$, $K = 1$, $T_{\min} = T_{\max}/2$, $T_{\max} = 10^4$, averages over $10^3$ realizations.
}
\label{fig:mean_population_sigma}
\end{figure}

We have explored the phase diagram of the stationary mean population density $x^\ast$ as a function of reproductive delay $\alpha$ and the noise amplitude $\sigma$ (see Figure~\ref{fig:mean_population_sigma}).

In the deterministic case ($\sigma=0$), the population reaches the stationary value $x^\ast = (1 - d/b)\,K$, which is independent of reproductive delay. Once environmental fluctuations are introduced ($\sigma>0$), the mean density becomes strongly delay-dependent: the largest values of $x^\ast$ occur for vanishing delays ($\alpha\to0$), and to a lesser extent for very prolongued delays. For intermediate $\alpha$, reproduction becomes significantly less effective, producing a pronounced depression in the population level. As $\sigma$ increases, this depression deepens and broadens, forming a valley of low mean population that may eventually drive the system towards extinction (see Figure~\ref{fig:mean_population_sigma}, lower panel).

A qualitatively similar diagram is obtained in the $(\alpha,\tau)$ plane for finite $\sigma$, where increasing $\tau$ has an effect analogous to increasing $\sigma$ (see Appendix \ref{appendix:meanpopulation-tau}). This will be further clarified in the following section.

\subsection{Mean growth rate}
To gain further insight into the aforementioned phenomenology, we analyze the linearized dynamics by taking the limit $K \rightarrow \infty$ in Eq. \ref{eq:model}. This effectively removes density-dependent saturation effects and isolates the influence of reproductive delay and environmental variability on the intrinsic growth dynamics:
\begin{equation}
\dot{x}(t) = (b + \sigma \xi_\tau(t))\, x(t-\alpha) - d\, x(t).
\label{eq:model-linear}
\end{equation}
In this regime, we compute the mean growth rate, defined as:
\begin{equation}
G= \lim_{T\rightarrow \infty} \left\langle\frac{\log\left(\frac{x(T)}{x(0)}\right)}{T} \right\rangle,
\end{equation}
where the average is taken over independent realizations of the environmental noise. In practice, the limit is approximated numerically by choosing a sufficiently large integration time, $T = T_\mathrm{max}$.

\begin{figure}
    \centering
    \includegraphics[width=\columnwidth]{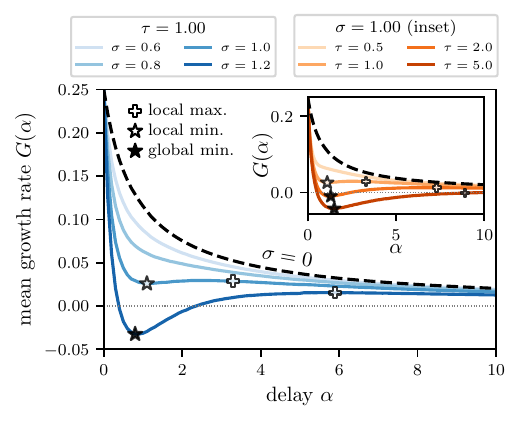}
    \caption{Mean growth rate $G$ as a function of reproductive delay $\alpha$, for different values of the noise amplitude $\sigma$ (main panel) and correlation time $\tau$ (inset), in the model with unbounded growth, $K=\infty$ (numerical results from simulations). While $G$ is maximal at $\alpha = 0$ for all cases, increasing $\sigma$ or $\tau$ leads to the emergence of a local minimum at intermediate delays that eventually becomes global (indicated with different symbols)}. The dashed lines correspond to the analytical solution of the deterministic case, $\sigma=0$ (see Eq.(\ref{eq:G0})). Parameters: $d = 1$, $b = 1.25$, $T_{\max}=10^3$, averages over $10^4$ independent realizations.
    \label{fig:galpha}
\end{figure}

The main panel of Figure~\ref{fig:galpha} shows the mean growth rate $G$ as a function of reproductive delay $\alpha$, for fixed correlation time $\tau$ and varying noise amplitude $\sigma$. The growth rate is always maximal at $\alpha=0$, which corresponds to the non-delayed reference model, i.e., the absence of an obligatory pre-reproductive interval. Its decay with increasing delays, however, depends strongly on the level of environmental variability. For small noise amplitudes, $G(\alpha)$ decreases monotonically. Beyond a certain threshold in $\sigma$, the curve develops a local minimum followed by a local maximum (i.e., via a fold, or saddle-node bifurcation in the language of dynamical systems~\cite{strogatz}). Notably, for sufficiently large noise amplitudes, this local minimum becomes global, making intermediate delays the least favorable strategy. This result is consistent with the previous section for finite $K$, and depressions in $G(\alpha)$ at intermediate delay durations mirror the valley of low stationary densities reported for $x^\ast$ (Figure~\ref{fig:mean_population_sigma}). Finally, in the asymptotic regime of prolonged delays, $\alpha\to\infty$, the influence of environmental variability vanishes and $G(\alpha)$ tends to the deterministic profile ($\sigma=0$).

The reported behavior becomes even more pronounced as the average birth rate $b$ approaches the death rate $d$ (see Appendix~\ref{appendix:b}), indicating that proximity to the extinction threshold magnifies the impact of reproductive delays and environmental correlations on $G$.

A similar qualitative behavior is observed when keeping $\sigma$ fixed and increasing the correlation time $\tau$ (see inset of Figure~\ref{fig:galpha}).

\subsubsection{Analytical calculation}
We develop a semi-analytical framework to further investigate the influence of each parameter and timescale on $G(\alpha)$. The literature provides general approaches to treat either stochastic differential equations with delay and white noise \cite{frank2005delay}, or non-delayed systems subject to colored noise \cite{UCNA}, but a unified framework that simultaneously accounts for both delay and time-correlated noise is extremely challenging and, to the best of our knowledge, still lacking. Here, we do not attempt to address this general problem, but rather aim to derive an approximate expression tailored to our model.
To this end, let us first analyze the deterministic case ($\sigma=0$), where the population grows exponentially with a growth rate $G_0$, $x(t)=x(0) e^{G_0 t}$. Introducing this expression into Eq.~\ref{eq:model-linear}, one finds that
\begin{equation}
G_0 = b e^{-\alpha G_0} - d = b_\mathrm{eff} -d,
\label{eq:G0}
\end{equation}
where we have introduced the effective birth rate $b_\mathrm{eff}=b e^{-\alpha G_0}$ for convenience. This implicit equation for $G_0(\alpha)$ is well known in the literature on delayed population models \cite{kuang1993delay} and must be solved numerically (dashed line in Figure~\ref{fig:galpha}).

It is instructive to analyze the behavior of $G_0(\alpha)$ at small and large delays. For short delays, expanding the exponential in Eq.~\eqref{eq:G0} to first order gives $G_0\simeq (b-d)/(1+\alpha b)\simeq (b-d)\,(1-\alpha b)$ (with $\alpha\ll b^{-1}$). This also shows that the initial decay scale of both $G_0$ and $b_\mathrm{eff}$ is set by $b^{-1}$. For large delays, setting $y=\alpha G_0$ shows that $y$ remains bounded and the balance $b e^{-y}\to d$ is required, hence $G_0(\alpha)\simeq \ln(b/d)/\alpha$.

In the presence of fluctuations, we can introduce the ansatz $x(t)=x(0) e^{-\int_{0}^t g(s) ds}$ in Eq.~\ref{eq:model-linear}, where $g(t)$ is the \textit{instantaneous} growth rate (note that $\lim_{t\rightarrow \infty}\langle g(t)\rangle=G$). We then obtain an equation for $g(t)$ that must be solved self-consistently:
\begin{equation}
g(t) = (b + \sigma \xi_\tau(t))\, e^{-\int_{t-\alpha}^t ds\, g(s)} - d.
\label{eq:g(t)}
\end{equation}
The integral can be estimated in two asymptotic limits.
For \textbf{small delays} ($\alpha \ll b^{-1}$), we approximate the integral by evaluating it at time $t$, $\int_{t-\alpha}^t g(s)\,ds \simeq \alpha g(t)$. Then we write $g(t) = G_0 + \sigma g_1(t)$ and Taylor expand $e^{-\alpha \sigma g_1(t)}$ to linear order in $\alpha$. We further assume that the noise amplitude $\sigma$ is small and discard all mixed terms of order $\sigma^2 \alpha$. Many terms cancel, yielding an explicit equation for $g_1(t)$:
\begin{equation}
g_1(t) = \frac{b_\mathrm{eff}/b}{1 + b_\mathrm{eff}\alpha}\, \xi_\tau(t).  
\end{equation}
This result can be introduced in Eq. \ref{eq:g(t)} to compute the mean growth rate $G$ for small delays:
{\footnotesize
\begin{align}
G_\mathrm{small} &= \left\langle 
(b + \sigma \xi_\tau(t))\, e^{-G_0 \alpha} 
\left(1 - \sigma \frac{b_\mathrm{eff}/b}{1 + b_\mathrm{eff} \alpha} 
\int_{t-\alpha}^t \xi_\tau(s)\, ds \right) 
\right\rangle \notag \\
&= G_0 - \sigma^2 \frac{(b_\mathrm{eff}/b)^2}{1 + b_\mathrm{eff}\alpha} 
\int_{t-\alpha}^t \langle \xi_\tau(t) \xi_\tau(s) \rangle\, ds \notag \\
&= G_0 - \frac{1}{2} \sigma^2 \tau 
\frac{(b_\mathrm{eff}/b)^2}{1 + b_\mathrm{eff} \alpha} \left(1 - e^{-2\alpha/\tau}\right).
\end{align}
}
where we have used $\langle \xi_\tau(t) \xi_\tau(s) \rangle = e^{-2|t - s|/\tau}$.

On the other hand, for \textbf{large delays} ($\alpha \gg b^{-1}$), the integral can be replaced by its mean value, $\int_{t-\alpha}^t g(s)\, ds \simeq \alpha G_\mathrm{large}$, and taking the average in Eq.~\ref{eq:g(t)} yields an implicit equation for $G_\mathrm{large}$ that coincides with Eq.~\ref{eq:G0}. Therefore,
\begin{equation}
G_\mathrm{large} = G_0,
\end{equation}
as indeed is observed in Figure~\ref{fig:galpha}.

We finally introduce a heuristic expression that interpolates between the two regimes, $
G(\alpha) = f(\alpha)\, G_\mathrm{small} + (1 - f(\alpha))\, G_\mathrm{large}$, 
where $f(\alpha)$ is an interpolation function,  e.g. $f(\alpha) = (1 + \alpha b)^{-1}$; 
other choices for this function yield similar performance and do not significantly affect the forthcoming results. Substituting the expressions, we obtain:
\begin{equation}
G = G_0 - \frac{1}{2} \sigma^2 \tau \frac{(b_\mathrm{eff}/b)^2}{(1 + b_\mathrm{eff} \alpha)(1+b\alpha)}  \left(1 - e^{-2\alpha/\tau}\right).
\label{eq:G-formula}
\end{equation}

\begin{figure}
    \includegraphics[width=\columnwidth]{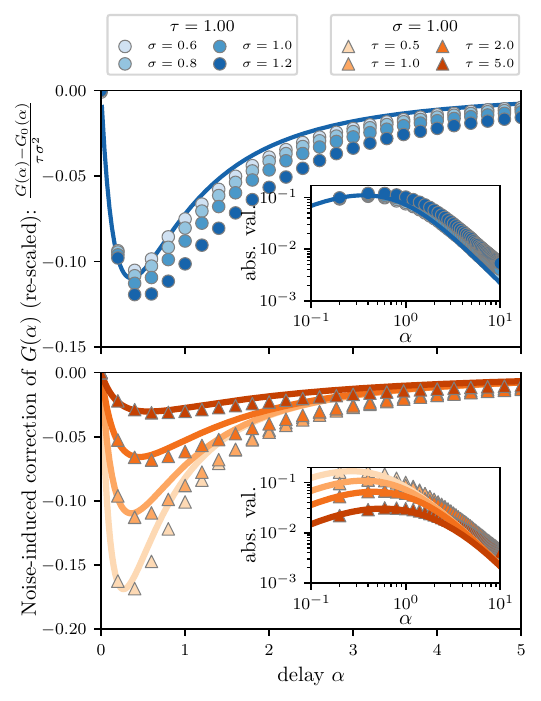}
    \caption{
    Noise-induced correction to the mean growth rate as a function of reproductive delay $\alpha$, rescaled by the factor $\tau \sigma^2$. Solid lines correspond to the approximation given in Eq.~\ref{eq:G-formula}, while points denote numerical simulation results. Top panel: fixed correlation time $\tau = 1$ and varying noise amplitude $\sigma$. Bottom panel: fixed $\sigma = 1$ and varying $\tau$. Insets show the same data (in absolute value) on log-log scale, highlighting the asymptotic decay at large values of $\alpha$. Parameters: $d = 1$, $b = 1.25$.}
    \label{fig:g_correction}
\end{figure}

We have assessed the validity of this approximation in Figure~\ref{fig:g_correction} (solid line), comparing it with numerical simulations for increasing values of $\sigma$ at fixed $\tau$ (upper panel), and increasing $\tau$ at fixed $\sigma$ (lower panel). Overall, the agreement is good: it captures the qualitative dynamics and the locations of the local extrema. This provides a semi-analytical expression that captures the dependence of the growth rate on the model parameters.

The noise-induced correction in Eq.~\ref{eq:G-formula} exhibits a single peak (see Figure~\ref{fig:g_correction}) from the competition between the decreasing term $\frac{(b_\mathrm{eff}(\alpha)/b)^2}{(1+\alpha\, b_\mathrm{eff}(\alpha))(1+b\alpha)}$ and the increasing factor $(1-e^{-2\alpha/\tau})$. Each factor acts on its own scale: the former decays over $b^{-1}$ (via $b_\mathrm{eff}$ and the denominators), while the latter grows over the environmental correlation time $\tau/2$. Although we cannot provide a closed-form expression for the peak location ---marking the time of worst performance--- its scale is set by the shorter limiting timescale, $\alpha_\ast \sim \min\{\tau/2,\,b^{-1}\}$.

\subsection{Extinction times}
We now explore the resilience of the population to external perturbations by analyzing how delay duration impacts the mean extinction time. It is important to note that in order to observe true extinction ($x=0$), demographic fluctuations must be included in Eq. \ref{eq:model}. In the absence of demographic noise, the absorbing state cannot be reached in finite time \cite{Nature}.

To overcome this limitation, we adopt the following standard approach: we introduce an artificial absorbing boundary at $x = 1/N$, effectively mimicking the population density corresponding to a single individual in a finite population of size $N$. 
Starting from a given initial history ---here $x(t)=K/2$ for $t\in[-\alpha,0]$, with results robust to other smooth histories--- we compute the mean first-passage time $\bar T$ to this boundary, which we use as a proxy for the mean extinction time.
 This procedure yields results consistent with more elaborate demographic simulations while remaining computationally efficient~\cite{Spanio2017,Hidalgo2017}.

\begin{figure}
    \centering
    \includegraphics[width=\columnwidth]{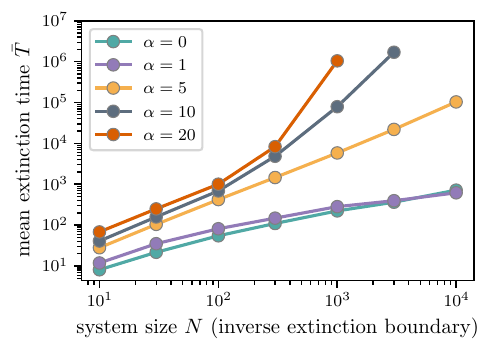}
    \caption{Mean extinction time $\bar{T}$ (measured as the time to reach the absorbing boundary $x = 1/N$) as a function of system size $N$, for different values of the reproductive delay $\alpha$. Parameters: $d = 1$, $b = 1.05$, $\tau = 1$, $\sigma = 0.5$, $K = 1$. Each point represents an average over $10^3$ realizations.}
    \label{fig:T}
\end{figure}

Figure \ref{fig:T} shows the dependence of $\bar{T}$ on the effective system size $N$ for different values of reproductive delay $\alpha$. In all cases, $\bar{T}$ increases with $N$. Interestingly, prolonged dormancy consistently leads to longer mean extinction times. For the set of parameters chosen here, distinct scaling regimes emerge depending on $\alpha$: $\bar{T}$ grows relatively slowly with $N$ when delays are short (i.e., sublinearly), indicating limited resilience. In contrast, for longer delays, the population becomes more stable, and $\bar{T}$ increases more rapidly with $N$. This behavior is consistent with the trends observed in Figure~\ref{fig:timeseries} (where we use the same parameter values to facilitate comparison), where large delays suppress both positive and negative fluctuations, thereby reducing extinction risk. This result highlights the stabilizing effect of reproductive delay under environmental variability.

\begin{figure}
    \centering
    \includegraphics[width=\columnwidth]{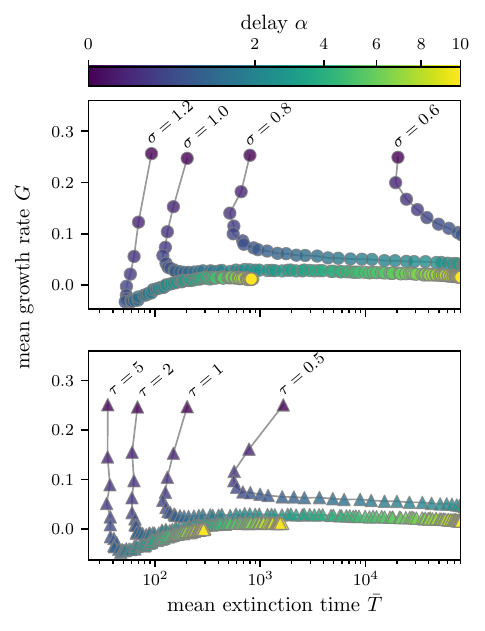}
    \caption{Trade-off between mean growth rate $G$ and mean extinction time $\bar{T}$ for different values of the reproductive delay $\alpha$. Each curve corresponds to a different level of environmental variability. Parameters: $d = 1$, $b= 1.25$, $\tau=1$ for variable $\sigma$ (top panel) and $\sigma=1$ for variable $\tau$ (bottom panel). To calculate $G$, we set $K= \infty$ and $T_{\max}=10^3$, and to calculate $\bar{T}$ we set $K=1$, $N=10^3$. In both cases, we average over $10^3$ realizations.}
    \label{fig:tradeoff}
\end{figure}

So far, we have analyzed two key indicators of population performance: the ability to grow under unbounded conditions, captured by the mean growth rate $G$ (in the limit $K \to \infty$), and the ability to persist over time, measured by the mean extinction time $\bar{T}$ (with an absorbing boundary at $x = N^{-1}$). In Figure~\ref{fig:tradeoff}, we make an explicit comparison between these two quantities both by fixing the correlation time $\tau$ and varying the noise amplitude $\sigma$ (upper panel), and by fixing $\sigma$ and varying $\tau$ (lower panel). Each point corresponds to a different value of reproductive delay $\alpha$.

As expected, the system faces a trade-off: short delays maximize growth ($G$) but results in shorter extinction times ($\bar{T}$), whereas prolonged lags promote resilience at the cost of slower growth. These differences become more pronounced under strong ($\sigma$ large) or persistent ($\tau$ large) environmental fluctuations. Interestingly, both panels reveal a non-monotonic behavior: the curve bends backward for intermediate values of $\alpha$, simultaneously lowering $G$ and $\bar{T}$. These regimes correspond to the least favorable strategies identified in previous sections. 

\section{Agent-based, evolutionary dynamics}
\label{sec:ABM}
 The results developed so far show that demographic delay interacting with temporally correlated environmental variability generates a fitness landscape with two adaptive peaks and a maladaptive intermediate region. However, these analyses quantify growth and persistence, not evolutionary success. In real biological systems, natural selection operates on populations composed of individuals with heritable variation, demographic stochasticity, and finite lifetimes. Whether the non-monotonic structure identified above truly translates into evolutionary dynamics is therefore not guaranteed.

To test whether the predicted bistability represents a genuine evolutionary mechanism, we developed an explicit agent-based model in which the reproductive delay is an evolving trait subject to mutation and natural selection (see lower panel of Figure \ref{fig:scheme}). The ecological setting matches the stochastic environment studied in Section \ref{sec:mean-field-model}: resources fluctuate between ``good'' and ``bad'' states according to a dichotomous Markov process with correlation time $\tau$. Agents consume resources, reproduce proportionally to intake, and generate propagules that remain inactive until they establish. Importantly, each individual carries an inheritable activation rate $\alpha$, which governs the probability per unit time that a propagule becomes active. While this formulation introduces natural variability in delay duration, its role here is not to modify the phenomenology of delay, but to allow delay strategies to evolve in a biologically realistic manner. Mortality affects only active agents, enabling dormant propagules to form a persistent reservoir analogous to seed banks or microbial persisters. All ecological and evolutionary events are simulated in continuous time using a Gillespie algorithm \cite{Gillespie}, ensuring full demographic and environmental stochasticity.

\begin{figure}
    \centering
    \includegraphics[width=\columnwidth]{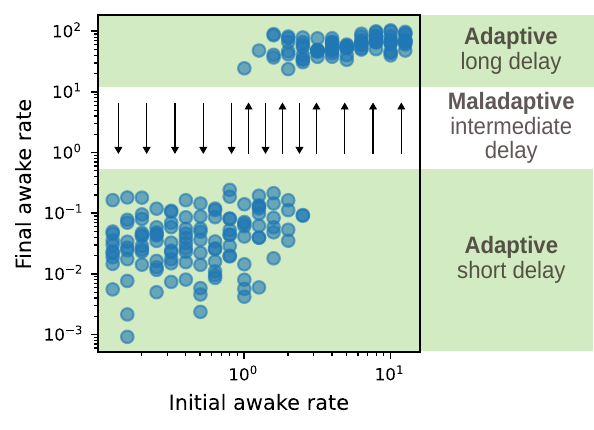}
    \caption{Evolutionary validation of the predicted bistable reproductive-delay strategies. Final population-mean delay rates $\alpha$ are plotted against their initial values across independent evolutionary simulations. The population consistently evolves toward one of two adaptive attractors: a short-delay strategy or a long-delay strategy, while avoiding intermediate values. Populations initiated in the maladaptive region never remain there; instead, natural selection drives them toward one of the two fitness peaks predicted by the delayed-logistic analysis. Parameter values: $\mu=100$, $\sigma=90$, $d=0.1$, $\nu=0.01$ and $\tau=4$; simulation time is $2\times10^5$.}
    \label{fig:agent-based-model}
\end{figure}

Despite these additional sources of variability and the lack of any imposed fixed delay, the evolutionary outcomes closely mirror the predictions derived from the delayed-logistic framework. As shown in Figure~\ref{fig:agent-based-model}, populations starting with low values of $\alpha$ evolve toward a stable short-delay strategy, while those starting with large $\alpha$ converge to a stable long-delay strategy. Crucially, populations initiated with intermediate $\alpha$ do not remain there. Instead, stochastic evolutionary trajectories drift away from the maladaptive region, ultimately reaching one of the two adaptive strategies. This confirms that the performance valleys and peaks identified in Section~\ref{sec:mean-field-model} represent evolutionary features of the fitness landscape generated by our model, rather than artifacts of fixed-delay assumptions or idealized population-level dynamics.

The emergence of evolutionary bistability in this fully stochastic model parallels empirical observations of timing strategies in nature, such as those found in microbial persisters, plant seed banks, and other systems where populations diversify into fast-activating and long-dormant phenotypes. These results demonstrate that the dual adaptive strategies revealed by our analysis are robust to demographic noise, architecture of the reproductive delay, and details of trait inheritance, reinforcing the view that the interaction between demographic memory and environmental temporal structure fundamentally shapes the evolution of reproductive maturation times.

\section{Conclusions and Discussion}
We have investigated how the duration of the reproductive delay shapes population performance in environments characterized by temporal correlations. Using a minimal delayed-logistic model with stochastic birth rates, we showed that delayed reproduction and colored environmental noise interact in a highly nontrivial way, producing distinct dynamical regimes and revealing a general principle: the adaptive value of reproductive delay depends critically on the relationship between the delay time and the environmental autocorrelation timescale.

Our analysis demonstrates that maturation time plays a dual, tightly balanced role as a reproductive delay. Very short delays maximize the growth rate, exploiting favorable conditions but amplifying susceptibility to rare downturns. Conversely, prolonged delays buffer environmental variability, enhancing resilience and substantially increasing extinction times even if at the cost of limited growth. Most strikingly, our work reveals the emergence of maladaptive intermediate delays —lag times that simultaneously reduce both growth and persistence below the values of both adaptive phases. This non-monotonous dependence on the reproductive delay arises robustly from the mismatch between demographic memory and the temporal structure of environmental variability, and it appears across numerical simulations, analytical approximations, and evolutionary models.

These findings show that the fitness landscape of reproductive maturation is not smooth but organized around a bistable structure, with two adaptive peaks separated by a broad maladaptive valley of intermediate delay. This bistability is confirmed in a fully stochastic, evolutionary agent-based model: populations consistently evolve toward either short or long reproductive delays, avoiding the intermediate region. Such evolutionary outcomes match empirical patterns observed in diverse biological systems, including bimodal germination strategies in plant seed banks ---where mixed populations of fast-germinating individuals and a persistent dormant reservoir coexist \cite{Kuipers,Childs,Toxin_Balaban,Tarnita2015,Tarnita2017,Loreau,Villa2}--- and phenotypic diversification into active and dormant subpopulations in microbes. It has also been posited that positively autocorrelated environments should select for long diapause in insects \cite{Hooper2025}.  In cancer, transitions between quiescent and proliferative states suggest that both extremes can contribute to the survival and spread of tumors. Our results can provide a theoretical framework to understand why intermediate, partially active states where cells are neither fully dormant nor fully proliferative, are particularly susceptible to therapeutic intervention \cite{cancer-Wang, cancer1, cancer2}. Further research should expand on this to clarify why intermediate or transitional dormancy stages are maladaptive and to what extent this can be manipulated by exogenous factors.

It is important to emphasize, however, that these biological connections should be understood as qualitative analogies rather than direct quantitative applications of the model. Our framework applies most directly to life histories that incorporate a clearly defined, unambiguous reproductive delay: individuals or propagules go through a lag phase before contributing to population growth, activation after this lag is not explicitly conditioned on environmental sensing, and environmental variability acts through the reproductive term. These assumptions leave out several mechanisms that are central to some specific biological systems. For instance, in microbial persistence, dormancy usually arises through phenotypic heterogeneity, with only a minority of cells entering low-activity persister states, rather than as an obligatory stage in the life cycle of all newborn cells \cite{Norman-review,Berkens}. In seed germination, exit from dormancy often depends on environmental cues whose reliability shapes germination timing and the evolution of bet-hedging strategies \cite{Donohue,Rubio2017}. In tumor progression, quiescent and proliferative states are shaped by cell-state heterogeneity and micro-environmental feedbacks, rather than by a uniform post-division delay imposed on every cell \cite{cancer1,cancer2}. Hence, our results should be interpreted as identifying a minimal mechanism by which reproductive delays and correlated environmental noise generate non-monotonic fitness landscapes, rather than as a full mechanistic model that can be applied directly to any of these systems.

Although our results may lack the necessary realism to fit specific biological examples, the qualitative patterns they describe are likely representative of many real-world scenarios. Reproductive delay is not a passive effect in life-history but an adaptive mechanism tuned to the timescales of environmental-variability autocorrelation. Consequently, organisms should evolve reproductive lag times that match the predictability of their environment. Other authors have shown empirically that dormancy can be regarded as insurance that organisms overcome the (potentially long) stretch of unfavorable conditions and that reproduction coincides with a favorable window (e.g., a specific seasonal cycle) \cite{Donohue, Kokko, Hooper2025}. Alternatively, selection could favor "r-strategists" that forego lags in reproductive maturation to ensure rapid exploitation of the environment, while it seems unlikely that intermediate strategies can persist in autocorrelated environments. We believe that this study provides a minimal but powerful theoretical foundation for understanding how environmental heterogeneity shapes adaptive reproductive delay strategies across biological domains. It opens the door to exploring evolutionarily stable distributions of lag times, multimodal bet-hedging strategies, and the ecological and physiological constraints that may shape the observed diversity of reproductive delay mechanisms in nature.

Beyond its biological implications, the present work underscores fundamental theoretical challenges. The model highlights the complexity of delayed, non-Markovian stochastic processes driven by correlated multiplicative noise near absorbing boundaries, an area where rigorous analytical tools remain limited. Our approximate analytical treatment provides initial insight, but future progress will require developing general methods to handle delay kernels, correlated noise, and extinction dynamics in a unified framework.

\section*{Data availability}
Simulation and analysis codes used in this work are openly available on Zenodo \cite{code}.

\section*{Acknowledgments}
We thank J. Grilli for his very insightful comments and suggestions. We are also extremely grateful to A. Maritan, S. Suweis, S. Azaele, M. Sireci and J.M. Camacho for enjoyable past collaborations on closely related topics.
This work has been supported by Grants Nos.  PID2023-149174NB-I00 and PID2022-143099OB-I00 financed by the Spanish Ministry and Agencia Estatal de Investigación MICIU/AEI/10.13039/501100011033 and European Regional Development Funds (ERDF). Thanks to the European Union Next-Generation EU (PIANO NAZIONALE DI RIPRESA E RESILIENZA (PNRR) – MISSIONE 4 COMPONENTE 2, “Dalla ricerca all'impresa” INVESTIMENTO 1.4 – D.D. 1034 17/06/2022, CN00000033).

\section*{Appendices}
\appendix
\section{Numerical integration}
\label{appendix:numericalintegration}

We use an explicit Runge-Kutta method of fourth order adapted to handle delay differential equations, following the procedure outlined in~\cite{Gros2019-reviewdelay}. The key idea is to discretize time with a fixed step $\Delta t$ and store the past history of the solution in a buffer or ring array, enabling efficient evaluation of the delayed term $x(t-\alpha)$ by interpolation. In our simulations, we typically use time steps $\Delta t \lesssim 0.01$, which ensures numerical stability and accuracy across the range of delays and noise intensities considered.

For the linearized dynamics, we perform the numerical integration after a change of variables. Since, in the absence of saturation, Eq.~\ref{eq:model-linear} exhibits unbounded exponential growth or decay, it is more natural to integrate it in logarithmic space. Such transformations have also been used in related problems to preserve positivity and avoid numerical artifacts in the direct integration \cite{mamunoz-PRL-logtransformation}. In particular, defining $z=\log x$, Eq.~\ref{eq:model-linear} becomes:

\begin{equation}
\label{eq:dynamics-log}
\dot z = (b+\sigma\,\xi_\tau(t))\, e^{z(t-\alpha)-z(t)} - d.
\end{equation}
No additional stochastic calculus (e.g., Itô corrections) is required, as the noise has finite correlation time.

\section{Dependence on the average birth rate $b$}
\label{appendix:b}

\begin{figure}
    \centering
    \includegraphics[width=\columnwidth]{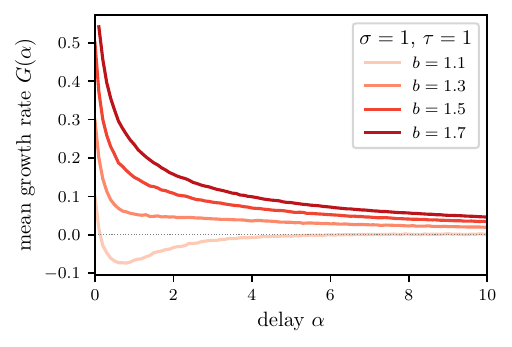}
\caption{Mean growth rate $G$ as a function of the reproductive delay $\alpha$ for different mean birth rates $b$ (simulations with the model presented in the main text). The non-monotonic behaviour of $G(\alpha)$ and the emergence of least favorable strategies become more evident when $b\rightarrow d$, approaching the extinction threshold. Parameters: $d=1$, $\sigma=1$, $\tau=1$, $T_{\max}=10^3$, averages over $10^2$ independent realizations.}
    \label{fig:birth}
\end{figure}

Figure \ref{fig:birth} shows the dependence of the mean growth rate $G(\alpha)$ on the reproductive delay $\alpha$ for different values of the mean birth rate $b$. 
As expected, the overall growth rate $G$ decreases as $b$ approaches the critical threshold $d$. More interestingly, the emergence of least favorable strategies—signaled by local minima in $G(\alpha)$—becomes more pronounced in this marginal regime. This reinforces the idea that proximity to the extinction threshold amplifies the influence of reproductive delay and environmental structure on population performance.

\section{Mean population diagram in the $(\alpha,\tau)$ plane}
\label{appendix:meanpopulation-tau}
\begin{figure}
    \centering
    \includegraphics[width=\columnwidth]{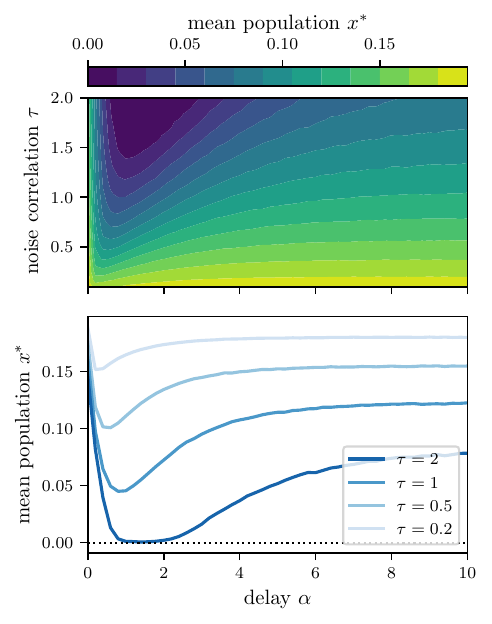}
    \caption{
    Mean population density $x^*$ as a function of reproductive delay $\alpha$ and noise correlation time $\tau$. Upper panel: heat map of $x^*$ in the $(\alpha,\tau)$ plane. Lower panel: the same data shown as $x^*$ versus $\alpha$ for different fixed values of $\tau$, highlighting its non-monotonic dependence. Increasing $\tau$ leads to a deepening and widening of the minimum at intermediate delays, analogous to the effect of increasing noise amplitude $\sigma$. Parameters: $d = 1$, $b = 1.25$, $\sigma = 1$, $K = 1$, $T_{\min} = T_{\max}/2$, $T_{\max} = 10^4$, averages over $10^3$ realizations.
    }
    \label{fig:mean_population_tau}
\end{figure}

We explore the stationary mean population density $x^\ast$ as a function of reproductive delay $\alpha$ and the noise correlation time $\tau$ (Figure~\ref{fig:mean_population_tau}). For finite noise amplitude ($\sigma=1$), increasing $\tau$ produces an effect qualitatively similar to increasing $\sigma$ in the main text: intermediate values of $\alpha$ lead to a marked reduction in $x^\ast$, generating a broad minimum in the population density. This confirms that both the intensity and the temporal correlations of environmental fluctuations shape population viability in a comparable way.

\section{Environment modeled as an Ornstein-Uhlenbeck process}
\label{appendix:OU}
The environment can be also modeled as an Ornstein-Uhlenbeck process: 
\begin{equation}
    \dot\xi_\mathrm{OU}(t)=-\frac{1}{\tau_\mathrm{OU}}\xi_\mathrm{OU} + \sqrt{\frac{2}{\tau_\mathrm{OU}}} \eta(t),
\end{equation}
where $\eta(t)$ is a zero mean, delta-correlated white Gaussian noise. In this case, the environment has zero mean and characteristic time of temporal correlation $\tau_\mathrm{OU}$ \cite{Gardiner}.

\begin{figure}
    \centering
\includegraphics[width=\columnwidth]{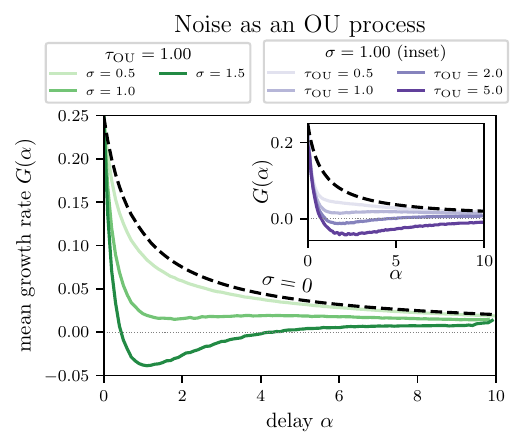}
\caption{Mean growth rate $G$ as a function of the reproductive delay $\alpha$ when the birth rate is driven by an Ornstein--Uhlenbeck process and rectified as $\max\!\big(0,\,b'+\sigma\,\xi_\mathrm{OU}(t)\big)$. Main panel: dependence on the noise amplitude $\sigma$ at fixed $\tau_\mathrm{OU}$; inset: dependence on the correlation time $\tau_\mathrm{OU}$ at fixed $\sigma$ (simulation results). In all cases, the same phenomenology emerges as for dichotomic noise (see main text). To enable a fair comparison, $b'$ is chosen so that the time-average of the birth rate equals $b$. Parameters: $d=1$, $b=1.25$; $b'(\sigma{=}0.5)=1.249$, $b'(\sigma{=}1)=1.194$, $b'(\sigma{=}1.5)=1.035$; $T_{\max}=10^3$; averages over $10^3$ independent realizations. }
    \label{fig:OU}
\end{figure}

As discussed in the main text, care is needed when integrating regimes where the effective birth rate can become negative: even the simple delay equation $\dot x=-x(t-\alpha)$ has the (a priori stable) fixed point $x^\ast=0$ losing stability beyond a critical delay, and the same mechanism appears in models with delayed birth and death terms \cite{Gros2019-reviewdelay,kuang1993delay}. For this reason, in simulations we enforce a non-negative birth rate to avoid unphysical destabilization induced by delay.

To address this issue, we model the birth rate as $\max\!\big(0,\, b' + \sigma\,\xi_{\mathrm{OU}}(t)\big)$ (i.e., no offspring are produced when the environmental variable falls below a threshold). However, when $\sigma \gtrsim b'$, the time-averaged birth rate $b$ differs from $b'$. To enable fair comparisons across $\sigma$, we therefore choose $b' = b'(\sigma)$ so that the average birth rate is fixed to $b$. Results are shown in Figure~\ref{fig:OU}. The same phenomenology as in the case studied in the main text (with dichotomic noise) emerges, highlighting the robustness of the observed behavior.

\section{Environmental variability in the death rate}
\label{appendix:d}

In the main text, we considered environmental variability acting on the birth process. Here, we explore an alternative scenario in which fluctuations instead affect the death rate. Since reproductive delay modulates only the birth term, the delay remains confined to reproduction, and the stochastic differential equation becomes:
\begin{equation}
\label{eq:dynamics_d}
\dot{x}(t) = b\, x(t-\alpha)\left(1 - \frac{x(t)}{K} \right) - \bigl(d + \sigma\, \xi_\tau(t)\bigr) x(t).
\end{equation}

\begin{figure}
    \centering
    \includegraphics[width=\columnwidth]{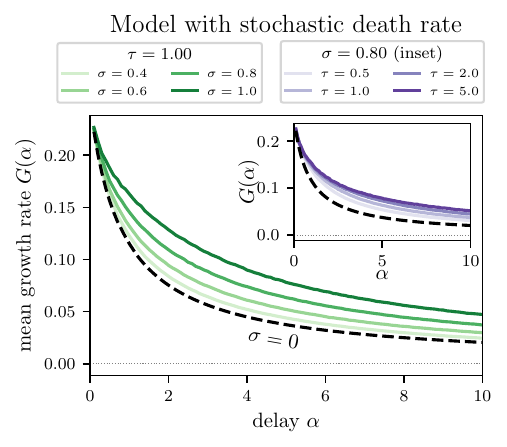}
\caption{Mean growth rate $G$ as a function of the reproductive delay $\alpha$, for different values of the noise amplitude $\sigma$ (main panel) and correlation time $\tau$ (inset), in the linearized model with stochastic death rate (simulation results). In contrast to the case with noise in the birth rate, here $G(\alpha)$ decreases monotonically with increasing delay, regardless of $\sigma$ or $\tau$. This highlights the crucial role of the coupling between noise and delay in generating non-monotonic behaviors. Parameters: $d = 1$, $b = 1.25$, $d=1$, $T_{\max}=10^3$, averages over $10^3$ independent realizations.}
    \label{fig:death}
\end{figure}

Figure~\ref{fig:death} shows the results of integrating this model in the linear regime ($K \to \infty$), using dichotomous Markov noise (DMN). In contrast to the case where fluctuations act on the birth rate, here the mean growth rate $G(\alpha)$ decays monotonically with the reproductive delay $\alpha$, without developing any local extrema. This indicates that the emergence of least favorable strategies—i.e., intermediate values of $\alpha$ that simultaneously reduce growth and persistence—is not a generic consequence of delay under noise. Instead, such regimes result specifically from the direct coupling between reproductive delay and environmental variability, which is absent when noise acts solely on the instantaneous death rate.

\newpage

\def\url#1{}
\bibliography{dormancy.bib}

@article{mamunoz-PRL-logtransformation,
  title={Systems with multiplicative noise: critical behavior from KPZ equation and numerics},
  author={Tu, Yuhai and Grinstein, G and Mu{\~n}oz, M.A.},
  journal={Physical review letters},
  volume={78},
  number={2},
  pages={274},
  year={1997},
  publisher={APS}
}

@book{Stearns1992,
  author    = {Stearns, Stephen C.},
  title     = {The Evolution of Life Histories},
  publisher = {Oxford University Press},
  address   = {Oxford},
  year      = {1992}
}

@article{Lloyd1966,
 ISSN = {00143820, 15585646},
 URL = {http://www.jstor.org/stable/2406568},
 author = {Monte Lloyd and Henry S. Dybas},
 journal = {Evolution},
 number = {2},
 pages = {133--149},
 publisher = {[Society for the Study of Evolution, Wiley]},
 title = {The Periodical Cicada Problem. I. Population Ecology},
 urldate = {2026-03-26},
 volume = {20},
 year = {1966}
}

@article{ODwyer2023,
  author  = {Jops, Kenneth and O'Dwyer, James P.},
  title   = {Life history complementarity and the maintenance of biodiversity},
  journal = {Nature},
  volume  = {618},
  pages   = {986--990},
  year    = {2023},
  doi     = {10.1038/s41586-023-06154-w}
}

@article{code,
  author  = {Hidalgo, Jorge and Fant, Lorenzo and Rubio de Casas, Rafael and Mu{\~n}oz, Miguel A.},
  title   = {Code for: Fluctuating environments favor extreme reproductive delays and penalize intermediate ones},
  journal = {Zenodo},
  year    = {2026},
  doi     = {10.5281/zenodo.17872776}
}

@book{strogatz,
  title={Nonlinear dynamics and chaos: with applications to physics, biology, chemistry, and engineering},
  author={Strogatz, Steven H},
  year={2024},
  publisher={Chapman and Hall/CRC}
}

@article{UCNA,
  title={Dynamical systems: a unified colored-noise approximation},
  author={Jung, Peter and H{\"a}nggi, Peter},
  journal={Physical review A},
  volume={35},
  number={10},
  pages={4464},
  year={1987},
  publisher={APS}
}

@article{frank2005delay,
  title={Delay Fokker-Planck equations, perturbation theory, and data analysis for nonlinear stochastic systems with time delays},
  author={Frank, TD},
  journal={Physical Review E},
  volume={71},
  number={3},
  pages={031106},
  year={2005},
  publisher={APS}
}

@article{Spanio2017,
  title={Impact of environmental colored noise in single-species population dynamics},
  author={Spanio, Tommaso and Hidalgo, Jorge and Mu{\~n}oz, Miguel A},
  journal={Physical Review E},
  volume={96},
  number={4},
  pages={042301},
  year={2017},
  publisher={APS}
}

@article{Hidalgo2017,
  title={Species coexistence in a neutral dynamics with environmental noise},
  author={Hidalgo, Jorge and Suweis, Samir and Maritan, Amos},
  journal={Journal of theoretical biology},
  volume={413},
  pages={1--10},
  year={2017},
  publisher={Elsevier}
}

@article{tuljapurkar1990,
  title={Delayed reproduction and fitness in variable environments.},
  author={Tuljapurkar, Shripad},
  journal={Proceedings of the National Academy of Sciences},
  volume={87},
  number={3},
  pages={1139--1143},
  year={1990}
}

@book{kuang1993delay,
  title={Delay differential equations},
  author={Kuang, Yang},
  year={1993},
  publisher={Academic press New York}
}

@article{Gros2019-reviewdelay,
  title={Chaos in time delay systems, an educational review},
  author={Wernecke, Hendrik and S{\'a}ndor, Bulcs{\'u} and Gros, Claudius},
  journal={Physics Reports},
  volume={824},
  pages={1--40},
  year={2019},
  publisher={Elsevier}
}

@article{Rubio2014,
title={The evolution of seed dormancy: environmental cues, evolutionary hubs, and diversification of the seed plants},
author={Willis, Charles G and Baskin, Carol C and Baskin, Jerry M and Auld, Josh R and Venable, D Lawrence and Cavender-Bares, Jeannine and Donohue, Kathleen and Rubio de Casas, Rafael and {{NGW Group}}},
journal={New Phytologist},
volume={203},
number={1},
pages={300--309},
year={2014},
publisher={Wiley Online Library}
}

@article{Rubio2017,
  title={Global biogeography of seed dormancy is determined by seasonality and seed size: a case study in the legumes},
  author={Rubio de Casas, Rafael and Willis, Charles G and Pearse, William D and Baskin, Carol C and Baskin, Jerry M and Cavender-Bares, Jeannine},
  journal={New Phytologist},
  volume={214},
  number={4},
  pages={1527--1536},
  year={2017},
  publisher={Wiley Online Library}
}

@article{Donohue,
  author  = {Donohue, Kathleen and Rubio de Casas, Rafael and Burghardt, Liana
             and Kovach, Katherine and Willis, Charles G.},
  title   = {Germination, Postgermination Adaptation, and Species Ecological Ranges},
  journal = {Annual Review of Ecology, Evolution, and Systematics},
  year    = {2010},
  volume  = {41},
  pages   = {293--319},
  doi     = {10.1146/annurev-ecolsys-102209-144715}
}

@article{LaFuerza-Toral2013,
  title={Stochastic description of delayed systems},
  author={Lafuerza, Luis F and Toral, Ra{\'u}l},
  journal={Philosophical Transactions of the Royal Society A: Mathematical, Physical and Engineering Sciences},
  volume={371},
  number={1999},
  pages={20120458},
  year={2013},
  publisher={The Royal Society Publishing}
}

@article{Bena2006,
  title={Dichotomous Markov noise: exact results for out-of-equilibrium systems},
  author={Bena, Ioana},
  journal={International Journal of Modern Physics B},
  volume={20},
  number={20},
  pages={2825--2888},
  year={2006},
  publisher={World Scientific}
}

@article{Loreau,
  title={Biodiversity as insurance: from concept to measurement and application},
  author={Loreau, Michel and Barbier, Matthieu and Filotas, Elise and Gravel, Dominique and Isbell, Forest and Miller, Steve J and Montoya, Jose M and Wang, Shaopeng and Aussenac, Rapha{\"e}l and Germain, Rachel and others},
  journal={Biological Reviews},
  volume={96},
  number={5},
  pages={2333--2354},
  year={2021},
  publisher={Wiley Online Library}
}

@article{Fridman,
	author = {Fridman, Ofer and Goldberg, Amir and Ronin, Irine and Shoresh, Noam and Balaban, Nathalie Q.},
	journal = {Nature},
	pages = {418--421},
	title = {Optimization of lag time underlies antibiotic tolerance in evolved bacterial populations},
	volume = {513},
	year = {2014}
}

@article{Lennon2011,
  title={Microbial seed banks: the ecological and evolutionary implications of dormancy},
  author={Lennon, Jay T and Jones, Stuart E},
  journal={Nature reviews microbiology},
  volume={9},
  number={2},
  pages={119--130},
  year={2011},
  publisher={Nature Publishing Group}
}

@article{Lennon2017,
  title={Evolution with a seed bank: the population genetic consequences of microbial dormancy},
  author={Shoemaker, William R and Lennon, Jay T},
  journal={Evolutionary applications},
  volume={11},
  number={1},
  pages={60--75},
  year={2018},
  publisher={Wiley Online Library}
}

@article{Nature,
  title={Nature of different types of absorbing states},
  author={Mu{\~n}oz, Miguel A},
  journal={Physical Review E},
  volume={57},
  number={2},
  pages={1377},
  year={1998},
  publisher={APS}
}

@article {Kussell-Leibler,
	author = {Kussell, Edo and Leibler, Stanislas},
	title = {Phenotypic Diversity, Population Growth, and Information in Fluctuating Environments},
		journal = {Science},
	volume = {309},
	number = {5743},
	pages = {2075--2078},
	year = {2005},
	publisher = {American Association for the Advancement of Science},
	issn = {0036-8075},
	URL = {https://science.sciencemag.org/content/309/5743/2075}
}

@article{Genovese,
  title={Recent results on multiplicative noise},
  author={Genovese, Walter and Mu\~noz, Miguel A},
  journal={Physical Review E},
  volume={60},
  number={1},
  pages={69},
  year={1999},
  publisher={APS}
}

@article{Villa2,
  title={Bet-hedging strategies in expanding populations},
  author={Villa Mart{\'\i}n, Paula and Mu{\~n}oz, Miguel A and Pigolotti, Simone},
  journal={PLoS computational biology},
  volume={15},
  number={4},
  pages={e1006529},
  year={2019},
  publisher={Public Library of Science}
}

@article{Rees,
  title={Evolutionary ecology of seed dormancy and seed size},
  author={Rees, Mark},
  journal={Philosophical Transactions of the Royal Society of London. Series B: Biological Sciences},
  volume={351},
  number={1345},
  pages={1299--1308},
  year={1996},
  publisher={The Royal Society London}
}

@article{Gillespie,
  title={A general method for numerically simulating the stochastic time evolution of coupled chemical reactions},
  author={Gillespie, Daniel T},
  journal={Journal of computational physics},
  volume={22},
  number={4},
  pages={403--434},
  year={1976},
  publisher={Elsevier Science}
}

@book{Gardiner,
  title={Stochastic Methods: A Handbook for the Natural and Social Sciences},
  author={Gardiner, C.},
  isbn={9783540707127},
  lccn={2008936877},
  series={Springer Series in Synergetics},
  year={2009},
  publisher={Springer}
}

@article {Toxin_Balaban,
	author = {Rotem, Eitan and Loinger, Adiel and Ronin, Irine and Levin-Reisman, Irit and Gabay, Chana and Shoresh, Noam and Biham, Ofer and Balaban, Nathalie Q.},
	title = {Regulation of phenotypic variability by a threshold-based mechanism underlies bacterial persistence},
	volume = {107},
	number = {28},
	pages = {12541--12546},
	year = {2010},
	publisher = {National Academy of Sciences},
	issn = {0027-8424},
	URL = {https://www.pnas.org/content/107/28/12541},
	journal = {Proceedings of the National Academy of Sciences}
}

@article{Kuipers,
author = {Veening, Jan-Willem and Smits, Wiep Klaas and Kuipers, Oscar P.},
title = {Bistability, Epigenetics, and Bet-Hedging in Bacteria},
journal = {Annual Review of Microbiology},
volume = {62},
number = {1},
pages = {193-210},
year = {2008},
    note ={PMID: 18537474},
URL = { 
        https://doi.org/10.1146/annurev.micro.62.081307.163002   
}
}

@article{Childs,
  title={Evolutionary bet-hedging in the real world: empirical evidence and challenges revealed by plants},
  author={Childs, Dylan Z and Metcalf, CJE and Rees, Mark},
  journal={Proceedings of the Royal Society B: Biological Sciences},
  volume={277},
  number={1697},
  pages={3055--3064},
  year={2010},
  publisher={The Royal Society}
}

@article{Mitarai,
  title={When to wake up? The optimal waking-up strategies for starvation-induced persistence},
  author={Himeoka, Yusuke and Mitarai, Namiko},
  journal={PLoS computational biology},
  volume={17},
  number={2},
  pages={e1008655},
  year={2021},
  publisher={Public Library of Science San Francisco, CA USA}
}

@article {Wide,
	author = {Moreno-G{\'a}mez, Stefany and Kiviet, Daniel J. and Vulin, Cl{\'e}ment and Schlegel, Susan and Schlegel, Kim and van Doorn, G. Sander and Ackermann, Martin},
	title = {Wide lag time distributions break a trade-off between reproduction and survival in bacteria},
	volume = {117},
	number = {31},
	pages = {18729--18736},
	year = {2020},
	doi = {10.1073/pnas.2003331117},
	publisher = {National Academy of Sciences},
	issn = {0027-8424},
	journal = {Proceedings of the National Academy of Sciences}
}

@article{powerlaw,
  title={Power-law tail in lag time distribution underlies bacterial persistence},
  author={{\c{S}}im{\c{s}}ek, Emrah and Kim, Minsu},
  journal={Proceedings of the National Academy of Sciences},
  volume={116},
  number={36},
  pages={17635--17640},
  year={2019},
  publisher={National Acad Sciences}
}

@article{Norman-review,
  title={Stochastic switching of cell fate in microbes},
  author={Norman, Thomas M and Lord, Nathan D and Paulsson, Johan and Losick, Richard},
  journal={Annual review of microbiology},
  volume={69},
  pages={381--403},
  year={2015},
  publisher={Annual Reviews}
}

@article{Baranyi,
  title={Stochastic modelling of bacterial lag phase},
  author={Baranyi, J{\'o}zsef},
  journal={International journal of food microbiology},
  volume={73},
  number={2-3},
  pages={203--206},
  year={2002},
  publisher={Elsevier}
}

@article{Venable,
  title={Delayed germination and dispersal in desert annuals: escape in space and time},
  author={Venable, D Lawrence and Lawlor, Lawrence},
  journal={Oecologia},
  volume={46},
  pages={272--282},
  year={1980},
  publisher={Springer}
}

@article{Lennon2021,
  title={Principles of seed banks and the emergence of complexity from dormancy},
  author={Lennon, Jay T and den Hollander, Frank and Wilke-Berenguer, Maite and Blath, Jochen},
  journal={Nature Communications},
  volume={12},
  number={1},
  pages={4807},
  year={2021},
  publisher={Nature Publishing Group UK London}
}

@article{Camacho,
  title={Phenotypic-dependent variability and the emergence of tolerance in bacterial populations},
  author={Camacho Mateu, Jos{\'e} and Sireci, Matteo and Mu{\~n}oz, Miguel A},
  journal={PLoS computational biology},
  volume={17},
  number={9},
  pages={e1009417},
  year={2021},
  publisher={Public Library of Science San Francisco, CA USA}
}

@article{Cohen1966,
  title={Optimizing reproduction in a randomly varying environment},
  author={Cohen, Dan},
  journal={Journal of theoretical biology},
  volume={12},
  number={1},
  pages={119--129},
  year={1966},
  publisher={Elsevier}
}

@article{Bulmer1984,
  title={Delayed germination of seeds: Cohen's model revisited},
  author={Bulmer, MG},
  journal={Theoretical Population Biology},
  volume={26},
  number={3},
  pages={367--377},
  year={1984},
  publisher={Elsevier}
}

@article{Lennon2023,
  title={Optimal dormancy strategies in fluctuating environments given delays in phenotypic switching},
  author={M{\u{a}}g{\u{a}}lie, Andreea and Schwartz, Daniel A and Lennon, Jay T and Weitz, Joshua S},
  journal={Journal of theoretical biology},
  volume={561},
  pages={111413},
  year={2023},
  publisher={Elsevier}
}

@article{color,
  title={The color of environmental noise},
  author={Vasseur, David A and Yodzis, Peter},
  journal={Ecology},
  volume={85},
  number={4},
  pages={1146--1152},
  year={2004},
  publisher={Wiley Online Library}
}

@book{Baskin2,
  title={Seeds: ecology, biogeography, and evolution of dormancy and germination},
  author={Baskin, Carol C and Baskin, Jerry M},
  year={2000},
  publisher={Academic press}
}

@article{Denlinger,
  title={Regulation of diapause},
  author={Denlinger, David L},
  journal={Annual review of entomology},
  volume={47},
  number={1},
  pages={93--122},
  year={2002},
  publisher={Annual Reviews 4139 El Camino Way, PO Box 10139, Palo Alto, CA 94303-0139, USA}
}

@article{Hutchinson1948,
  title={Circular causal systems in ecology},
  author={Hutchinson, G Evelyn and others},
  journal={Ann. NY Acad. Sci},
  volume={50},
  number={4},
  pages={221--246},
  year={1948}
}

@article{Lewontin1969,
  title={On population growth in a randomly varying environment},
  author={Lewontin, Richard C and Cohen, Daniel},
  journal={Proceedings of the National Academy of sciences},
  volume={62},
  number={4},
  pages={1056--1060},
  year={1969}
}

@book{Tuljapurkar,
  title={Population dynamics in variable environments},
  author={Tuljapurkar, Shripad},
  volume={85},
  year={2013},
  publisher={Springer Science \& Business Media}
}

@article{cancer1,
  title={Models, mechanisms and clinical evidence for cancer dormancy},
  author={Aguirre-Ghiso, Julio A},
  journal={Nature Reviews Cancer},
  volume={7},
  number={11},
  pages={834--846},
  year={2007},
  publisher={Nature Publishing Group UK London}
}

@article{cancer-Wang,
title = {Cancer dormancy and metabolism: From molecular insights to translational opportunities},
journal = {Cancer Letters},
volume = {635},
pages = {218097},
year = {2025},
issn = {0304-3835},
doi = {https://doi.org/10.1016/j.canlet.2025.218097},
url = {https://www.sciencedirect.com/science/article/pii/S030438352500669X},
author = {Yashi Wang and Lingyue Liu and Xiaozhen Zhang and Tingbo Liang and Xueli Bai}
}

@article{cancer2,
  title={The dormant cancer cell life cycle},
  author={Phan, Tri Giang and Croucher, Peter I},
  journal={Nature Reviews Cancer},
  volume={20},
  number={7},
  pages={398--411},
  year={2020},
  publisher={Nature Publishing Group UK London}
}

@article{Ratcliff,
  title={Bacterial persistence and bet hedging in Sinorhizobium meliloti},
  author={Ratcliff, William C and Denison, R Ford},
  journal={Communicative \& integrative biology},
  volume={4},
  number={1},
  pages={98--100},
  year={2011},
  publisher={Taylor \& Francis}
}

@article{China,
  title={Soil salinity regulates spatial-temporal heterogeneity of seed germination and seedbank persistence of an annual diaspore-trimorphic halophyte in northern China},
  author={Wang, Zhaoren and Baskin, Jerry M and Baskin, Carol C and Liu, Guofang and Ye, Xuehua and Yang, Xuejun and Huang, Zhenying},
  journal={BMC Plant Biology},
  volume={24},
  number={1},
  pages={604},
  year={2024},
  publisher={Springer}
}

@article{Cao,
  title={Comparison of germination and seed bank dynamics of dimorphic seeds of the cold desert halophyte Suaeda corniculata subsp. mongolica},
  author={Cao, Dechang and Baskin, Carol C and Baskin, Jerry M and Yang, Fan and Huang, Zhenying},
  journal={Annals of Botany},
  volume={110},
  number={8},
  pages={1545--1558},
  year={2012},
  publisher={Oxford University Press}
}

@article{Berkens,
  title={Integrative biology of persister cell formation: molecular circuitry, phenotypic diversification and fitness effects},
  author={Berkvens, Alicia and Chauhan, Priyanka and Bruggeman, Frank J},
  journal={Journal of the Royal Society Interface},
  volume={19},
  number={194},
  pages={20220129},
  year={2022},
  publisher={The Royal Society}
}

@article{Caceres,
  title={How long to rest: the ecology of optimal dormancy and environmental constraint},
  author={C{\'a}ceres, Carla E and Tessier, Alan J},
  journal={Ecology},
  volume={84},
  number={5},
  pages={1189--1198},
  year={2003},
  publisher={Wiley Online Library}
}

@article{Kokko,
  title={Optimal germination timing in unpredictable environments: the importance of dormancy for both among-and within-season variation},
  author={Ten Brink, Hanna and Gremer, Jennifer R and Kokko, Hanna},
  journal={Ecology letters},
  volume={23},
  number={4},
  pages={620--630},
  year={2020},
  publisher={Wiley Online Library}
}

@article{Tarnita2017,
  author  = {Ricardo Mart{\'i}nez-Garc{\'i}a and Corina E. Tarnita},
  title   = {Seasonality can induce coexistence of multiple bet-hedging strategies in \textit{Dictyostelium discoideum} via storage effect},
  journal = {Journal of Theoretical Biology},
  year    = {2017},
  volume  = {426},
  pages   = {104--116},
  doi     = {10.1016/j.jtbi.2017.05.019}
}

@article{Tarnita2015,
  author  = {Corina E. Tarnita and Alex Washburne and Ricardo Mart{\'i}nez-Garc{\'i}a and Allyson E. Sgro and Simon A. Levin},
  title   = {Fitness tradeoffs between spores and nonaggregating cells can explain the coexistence of diverse genotypes in cellular slime molds},
  journal = {Proceedings of the National Academy of Sciences},
  year    = {2015},
  volume  = {112},
  number  = {9},
  pages   = {2776--2781},
  doi     = {10.1073/pnas.1424242112}
}

@article{Levin1995,
  title={The timing of life history events},
  author={Iwasa, Yoh and Levin, Simon A},
  journal={Journal of Theoretical Biology},
  volume={172},
  number={1},
  pages={33--42},
  year={1995},
  publisher={Elsevier}
}

@article{HORN2012305,
title = {Phylogenetics and the evolution of major structural characters in the giant genus Euphorbia L. (Euphorbiaceae)},
journal = {Molecular Phylogenetics and Evolution},
volume = {63},
number = {2},
pages = {305-326},
year = {2012},
issn = {1055-7903},
doi = {https://doi.org/10.1016/j.ympev.2011.12.022},
url = {https://www.sciencedirect.com/science/article/pii/S1055790311005483},
author = {James W. Horn and Benjamin W. {van Ee} and Jeffery J. Morawetz and Ricarda Riina and Victor W. Steinmann and Paul E. Berry and Kenneth J. Wurdack}
}

@article{LeeOhSon2024EuphorbiaPlastome,
  author  = {Lee, Soo-Rang and Oh, Ami and Son, Dong Chan},
  title   = {Characterization, comparison, and phylogenetic analyses of chloroplast genomes of \textit{Euphorbia} species},
  journal = {Scientific Reports},
  volume  = {14},
  pages   = {15352},
  year    = {2024},
  doi     = {10.1038/s41598-024-66102-0},
  url     = {https://www.nature.com/articles/s41598-024-66102-0}
}

@article{SachidanandhamGin2009Dormancy,
  author  = {Sachidanandham, Ramasamy and Gin, K. Yew-Hoong},
  title   = {A dormancy state in nonspore-forming bacteria},
  journal = {Applied Microbiology and Biotechnology},
  volume  = {81},
  number  = {5},
  pages   = {927--941},
  year    = {2009},
  doi     = {10.1007/s00253-008-1712-y},
  url     = {https://doi.org/10.1007/s00253-008-1712-y}
}

@article{McDonaldDormancy2024,
Author = {Mcdonald, Mark D. and Owusu-Ansah, Carlos and Ellenbogen, Jared B. and
   Malone, Zachary D. and Ricketts, Michael P. and Frolking, Steve E. and
   Ernakovich, Jessica Gilman and Ibba, Michael and Bagby, Sarah C. and
   Weissman, J. L.},
Title = {What is microbial dormancy?},
Journal = {TRENDS IN MICROBIOLOGY},
Year = {2024},
Volume = {32},
Number = {2},
Pages = {142-150},
Month = {FEB},
DOI = {10.1016/j.tim.2023.08.006},
}

@article{Sweet2025,
title={Using diapause as a platform to understand the biology of dormancy},
author={Nathaniel A. Sweet and Chi-Kuo Hu},
journal={Open Biology},
year={2025},
volume={15},
doi={10.1098/rsob.250104}
}

@article{Wilsterman2020,
title={A unifying, eco‐physiological framework for animal dormancy},
author={Kathryn Wilsterman and M. Ballinger and C. Williams},
journal={Functional Ecology},
year={2020},
doi={10.1111/1365-2435.13718}
}

@article{Hooper2025,
title={Can long‐term diapause/dormancy improve persistence in a positively autocorrelated environment?},
author={Matthew Hooper and Jamie Musgrove and Francis Gilbert},
journal={Ecological Entomology},
year={2025},
volume={50},
pages={717 - 728},
doi={10.1111/een.13435}}
\end{document}